\documentclass[10pt]{iopart}

\usepackage[utf8]{inputenc}

\usepackage{siunitx}

\usepackage{amssymb}

\usepackage{hyphenat}

\usepackage{cite}

\usepackage{graphicx}

\usepackage{subfig}

\usepackage{stackengine}

\usepackage[export]{adjustbox}

\begin{document}

\title{Nanowire single-photon detectors made of atomic layer-deposited niobium nitride}

\author{E.~Knehr$^{1,2}$, A.~Kuzmin$^2$, M.~Ziegler$^1$, S.~Doerner$^2$, K.~Ilin$^2$, M.~Siegel$^2$, R.~Stolz$^1$ and H.~Schmidt$^{1,3}$}

\address{
	$^1$ Department of Quantum Detection, Leibniz Institute of Photonic Technology (Leibniz IPHT), Jena, Germany
	
	$^2$ Institute of Micro- and Nanoelectronic Systems, Karlsruhe Institute of Technology (KIT), Karlsruhe, Germany
	
	$^3$ Institute for Solid State Physics, University of Jena, Jena, Germany}
	
\ead{emanuel.knehr@leibniz-ipht.de}
\vspace{10pt}

\begin{abstract}
We demonstrate and characterize first superconducting nanowire single-photon detectors (SNSPDs) made from atomic layer- deposited (ALD) NbN layers.
To assess the suitability of these films as a detector material, transport properties of bare films and bridges of different dimensions and thicknesses are investigated.
Similar ratios of the measured critical current to the depairing current are obtained for micro-bridges made from ALD and sputtered NbN films.
Furthermore, we characterized the single-photon response for 5 and \SI{10}{nm}-thick nanowire detectors.
A \SI{100}{nm}-wide straight nanowire with a length of \SI{5}{\micro\meter} exhibits saturated count-rate dependencies on bias current and a cut-off wavelength in the near-infrared range.
The ALD technique could open up the possibility to fabricate NbN-based detectors on the wafer scale and to conformally cover also non-planar surfaces for novel device concepts.
\end{abstract}

\vspace{2pc}

\ioptwocol 

\section{Introduction}
Being developed since their introduction in 2001~\cite{Goltsman:2001}, superconducting nanowire single-photon detectors (SNSPDs) are very promising devices for applications in fields such as quantum cryptography, spectroscopy, and deep space communication~\cite{Takesue:2007,Natarajan:2012,Toussaint:2015}.

Key properties of SNSPDs include the spectral detection efficiency and the timing resolution limited by the timing jitter. By means of material research and optimization of the detector geometry and readout electronics, detection efficiencies of over \SI{90}{\percent} at \SI{1550}{nm} and a timing resolution as low as \SI{3}{ps} have been shown recently~\cite{Marsili:2013,Korzh:2018}. This sets SNSPDs apart from competing single-photon detectors (see table~1.1 in~\cite{Migdall:2013}).

For the use in SNSPDs, materials with a low superconducting energy gap, a small diffusion coefficient of the quasi-particles, and a critical current of nanowires close to the depairing current are advantageous~\cite{Semenov:2005,Dorenbos:2011a}. Especially the latter is necessary for a longer cut-off wavelength, but is often limited by inhomogeneities and constrictions in the active area of the detector~\cite{Kerman:2007,Charaev:2017}. Also, arrays of single-photon detectors require large-area ultra-thin films with a high uniformity with respect to the superconducting order parameter.

Besides recently emerging materials such as WSi~\cite{Baek:2011} and MoSi~\cite{Verma:2015}, NbN is still one of the most used materials because of its relatively high critical current densities and the established deposition technology.

So far, NbN-based SNSPDs are predominantly fabricated by reactive-magnetron sputtering of niobium in a nitrogen atmosphere. However, this technology cannot ensure high uniformity and homogeneity of the NbN films and is hard to apply for larger wafer sizes. Atomic layer deposition (ALD), on the other hand, could alleviate these problems by its chemical nature, depositing superconducting thin films monolayer by monolayer while making lower demands on the vacuum system and the temperature homogeneity of the substrate. Furthermore, other features of this deposition technique like the conformal coverage of pre-patterned substrates could lead to new applications.

Previously, ALD TiN films were reported for the application in SNSPDs~\cite{Morozov:2018}. In our earlier work, the deposition of superconducting NbN thin films using metal-organic plasma-enhanced atomic layer deposition (PEALD) has been shown~\cite{Ziegler:2013,Ziegler:2017,Linzen:2017}. Here, we investigate the suitability of these films for detector applications and demonstrate first SNSPDs and their properties.

\section{Fabrication technology}
We successfully deposited superconducting NbN in an Oxford Plasma Technology OpAL ALD system using (tert\hyp butylimido)\hyp tris(diethylamino)\hyp niobium (TBTDEN) and hydrogen plasma as precursors. The substrates are alternately exposed to these precursors to deposit monolayers of TBTDEN on the surface and reduce the monolayers to form NbN, respectively. In order to prevent gaseous phase reactions and limit the process to surface reactions, purge steps with argon as an inert gas are introduced between the precursor steps. By using a nitrogen glove box for sample handling, oxygen contaminations in the reactor could be minimized~\cite{Ziegler:2017}. The best film quality was achieved at a substrate temperature of \SI{350}{\celsius}, a plasma power of \SI{300}{W}, and a hydrogen flow rate of \SI{2.5}{sccm}. The resulting film growth per cycle was \SI{0.46}{\angstrom}. More details on the deposition process can be found in~\cite{Ziegler:2013,Ziegler:2017}. The NbN films were deposited on \SI{10x10}{mm} R-plane, double-side polished sapphire substrates and subsequently spin-coated with \SI{90}{nm} of poly-methyl-methacrylate (PMMA, 950~k).

Using electron-beam lithography, the resist was patterned in several steps. First, the nanowires were patterned in a negative-PMMA process similar to the process shown in~\cite{Charaev:2017}. After removing the unexposed areas with acetone, a new layer of PMMA was spun on and exposed in a positive process to pattern the readout lines, in which the detector is embedded. In a single step, the samples were etched via Ar-ion milling. After etching, the resist was again removed by acetone and 2-propanol.

The magnetron sputtering process, used to prepare NbN bridges for comparison (see section~3.2), is described in~\cite{Henrich:2012}.

\section{Transport measurements}

The bare films and test bridges were characterized by standard four-point measurements in a dipstick cooled down to \SI{4.2}{K} using liquid helium (lHe). For magnetic field measurements, a pulse tube cryocooler equipped with a superconducting magnet was used. The coils create a magnetic field up to \SI{5}{T} perpendicular to the substrate surface of the samples.

\subsection{NbN films}

The basic electrical characteristics of bare ALD-NbN films were investigated on samples with a thickness between 3 and \SI{21}{nm} (\SI{+-0.2}{nm}).

\begin{figure*}
	\centering
	\hspace{0pt}
	\subfloat{\includegraphics[width=0.49\textwidth,valign=t]{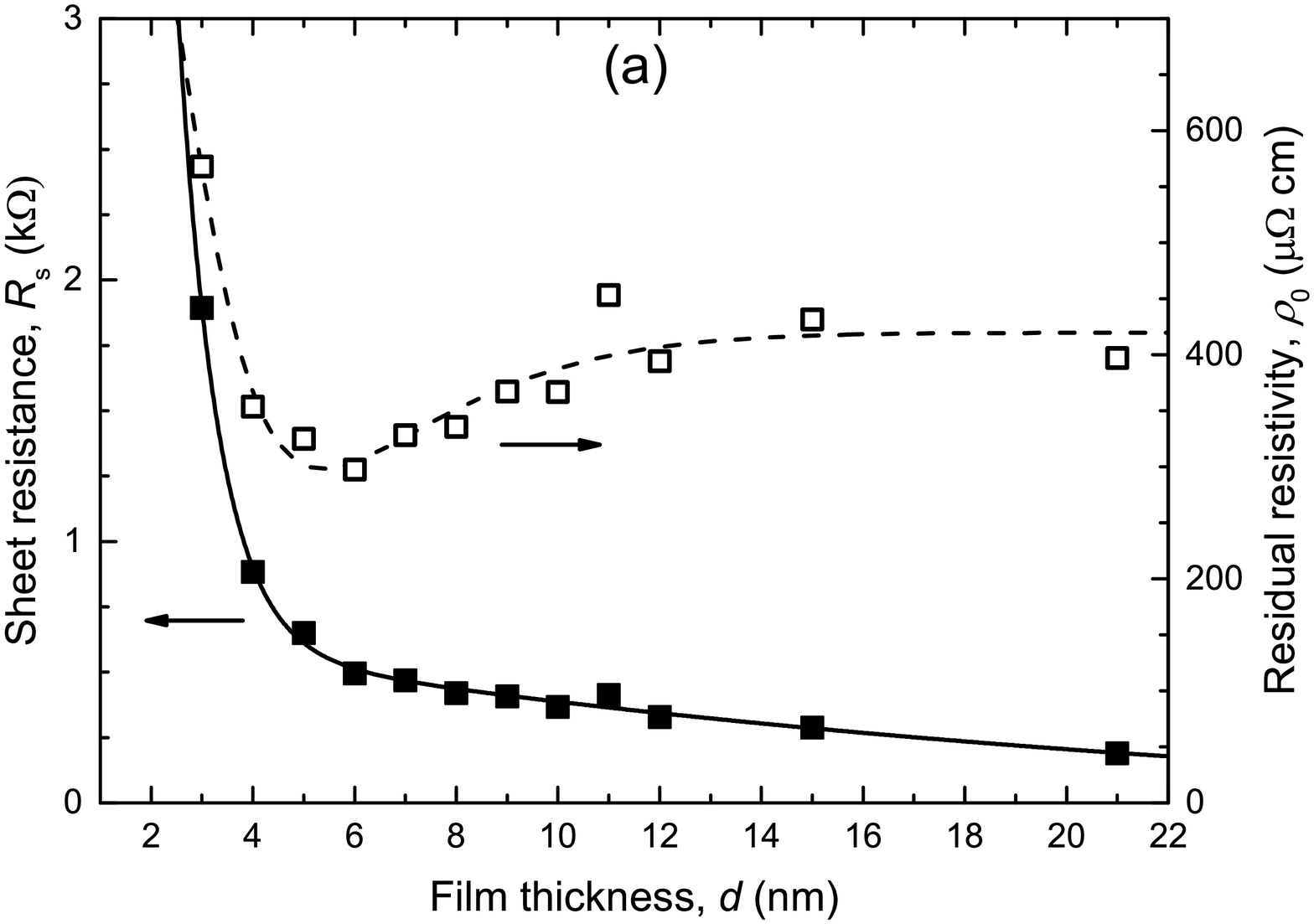}} \hfill
	\subfloat{\includegraphics[width=0.48\textwidth,valign=t]{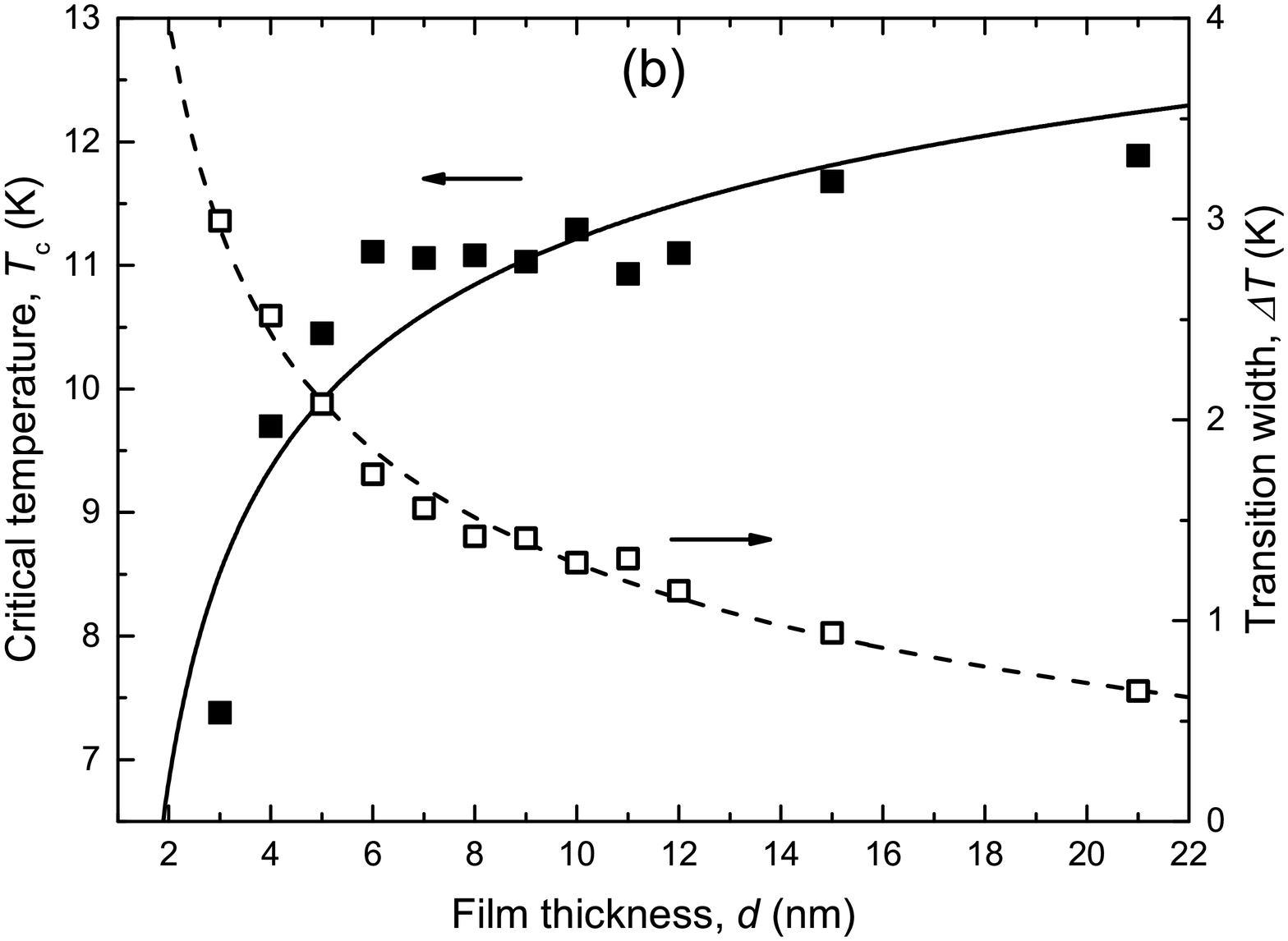}} \hfill
	\subfloat{\includegraphics[width=0.448\textwidth,valign=b]{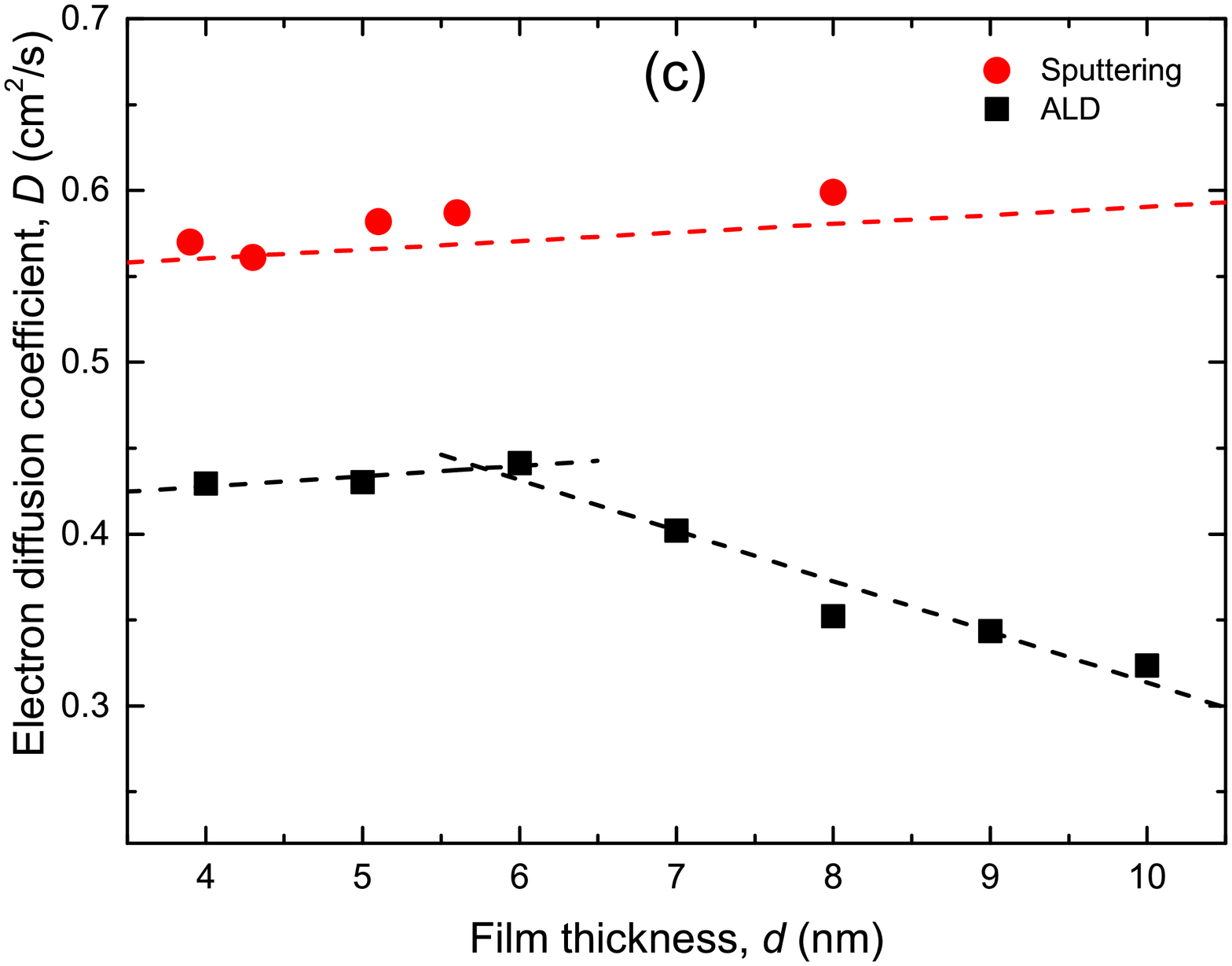}} \hfill
	\subfloat{\includegraphics[width=0.45\textwidth,valign=b]{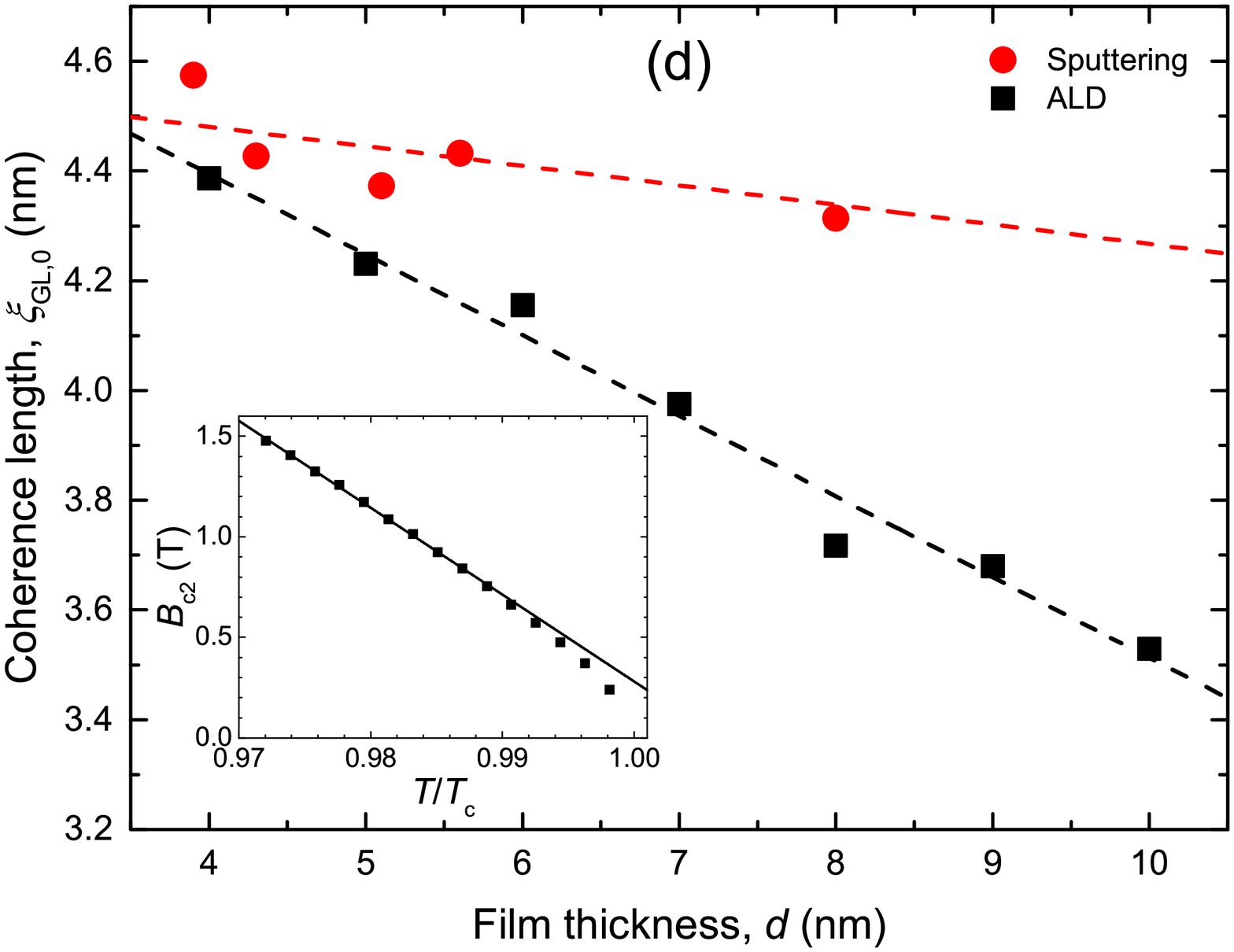}} \hspace{15pt}
	\caption{Electrical properties of ALD-NbN films with varying thickness; the sheet resistance $R_\mathrm{s}$ and residual resistivity $\rho_0$ in normal state (a), the critical temperature $T_\mathrm{c}$ and the transition width $\Delta T$ (b), the electron diffusion coefficient $D$ (c), and the Ginzburg-Landau coherence length at zero temperature $\xi_\mathrm{GL}(0)$ (d). The solid and dashed lines are to guide the eye. The comparison values for sputtered films are taken from~\cite{Semenov:2009}. (Inset) Temperature dependence of the upper critical magnetic field. The line represents a linear fit excluding the data points above $0.99\,T/T_\mathrm{c}$.}
	\label{fig:fig1}
\end{figure*}

In figure~1(a), the sheet resistance $R_\mathrm{s}$ in the normal state near the superconducting transition is plotted as a function of the thickness $d$. In addition, the residual resistivity $\rho_0 = R_\mathrm{s}d$ is shown. The lowest value for $\rho_0$ of about \SI{300}{\micro\ohm\,cm} corresponds to a film thickness of \SI{6}{nm}. For film thicknesses larger than \SI{10}{nm}, $\rho_0$ is almost constant, while it rapidly increases for $d < \SI{4}{nm}$.

In figure 1(b), both the critical temperature $T_\mathrm{c}$ (taken at $0.01\,R_\mathrm{s}$) and the width of the transition $\Delta T$ (temperature difference of $0.9$ and $0.1\,R_\mathrm{s}$) between the superconducting and the normal state are plotted over thickness $d$. Similar to the trend of $R_\mathrm{s}$, the critical temperature $T_\mathrm{c}$ reaches a plateau above $d = \SI{6}{nm}$. A value of $T_\mathrm{c} = \SI{11.3}{K}$ for a \SI{6}{nm} thick film has been obtained, while the critical temperature decreases rapidly for smaller film thicknesses, reaching \SI{7.4}{K} at \SI{3}{nm}.

The magnetic penetration depth $\lambda$ for \num{5} and \SI{10}{nm}-thick films, estimated from $\mu_0 \lambda ^2 = \hbar \rho_0 / \pi \beta_0 k_\mathrm{B} T_\mathrm{c}$ ($\beta_0 = 2.05$ for NbN~\cite{Pracht:2013}), amounts to about \SI{550}{nm}.

\begin{figure*}[t]
\centering
\begin{minipage}[t]{.485\textwidth}
	\centering
	\includegraphics[width=\linewidth]{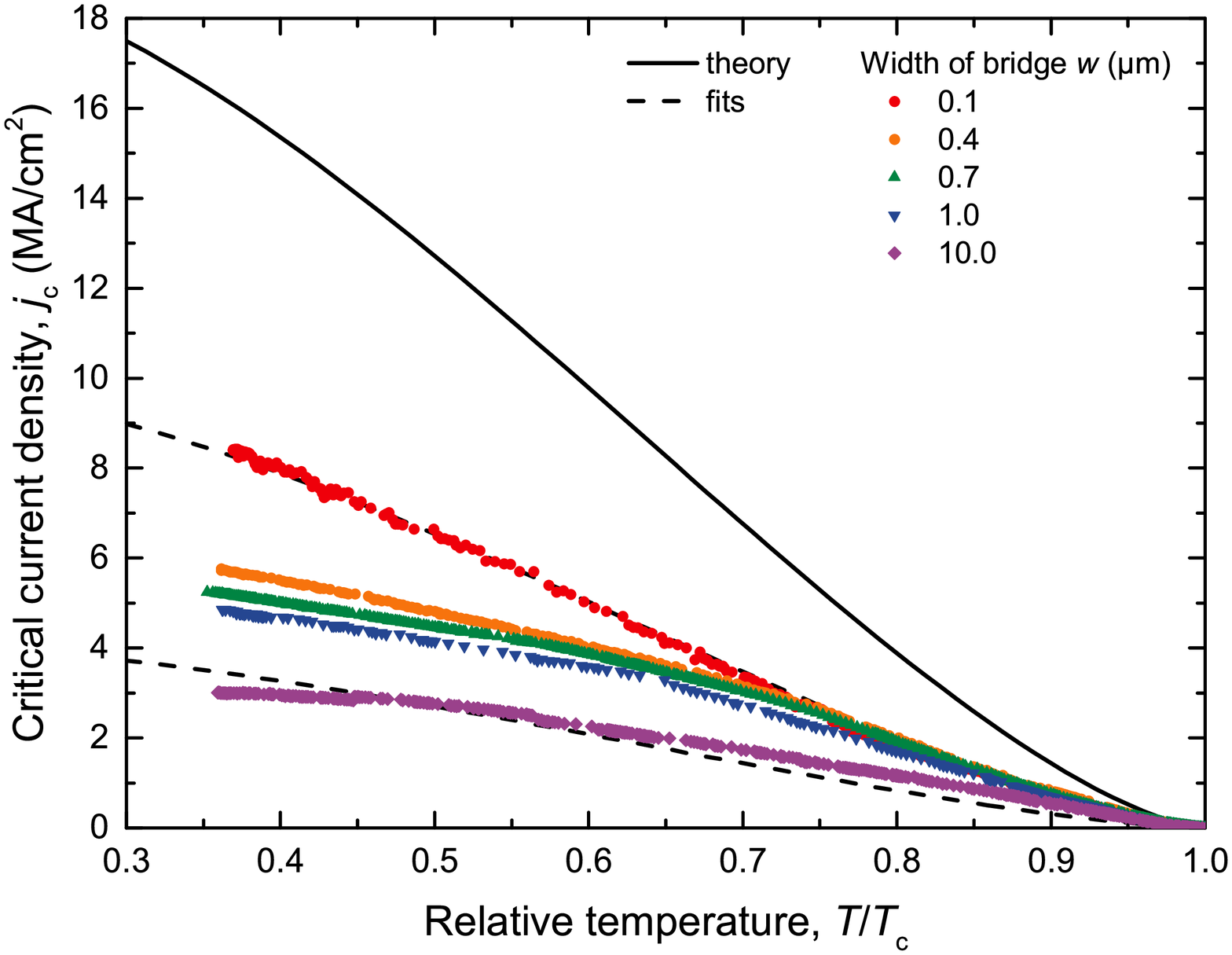}
	\captionof{figure}{Dependence of the critical current density $j_\mathrm{c}$ on temperature $T$ normalized to the critical temperature $T_\mathrm{c}$ for different bridge widths and $d_\mathrm{NbN} = \SI{10}{nm}$. The black line represents the temperature dependence of $j_\mathrm{dep}$ according to equation (4) and (5), while the dashed lines are fits to the data of the smallest and widest bridge with $j_\mathrm{dep,GL}(0)$ as fitting parameter.}
	\label{fig:fig2}
\end{minipage}\hfill%
\begin{minipage}[t]{.485\textwidth}
	\centering
	\includegraphics[width=\linewidth]{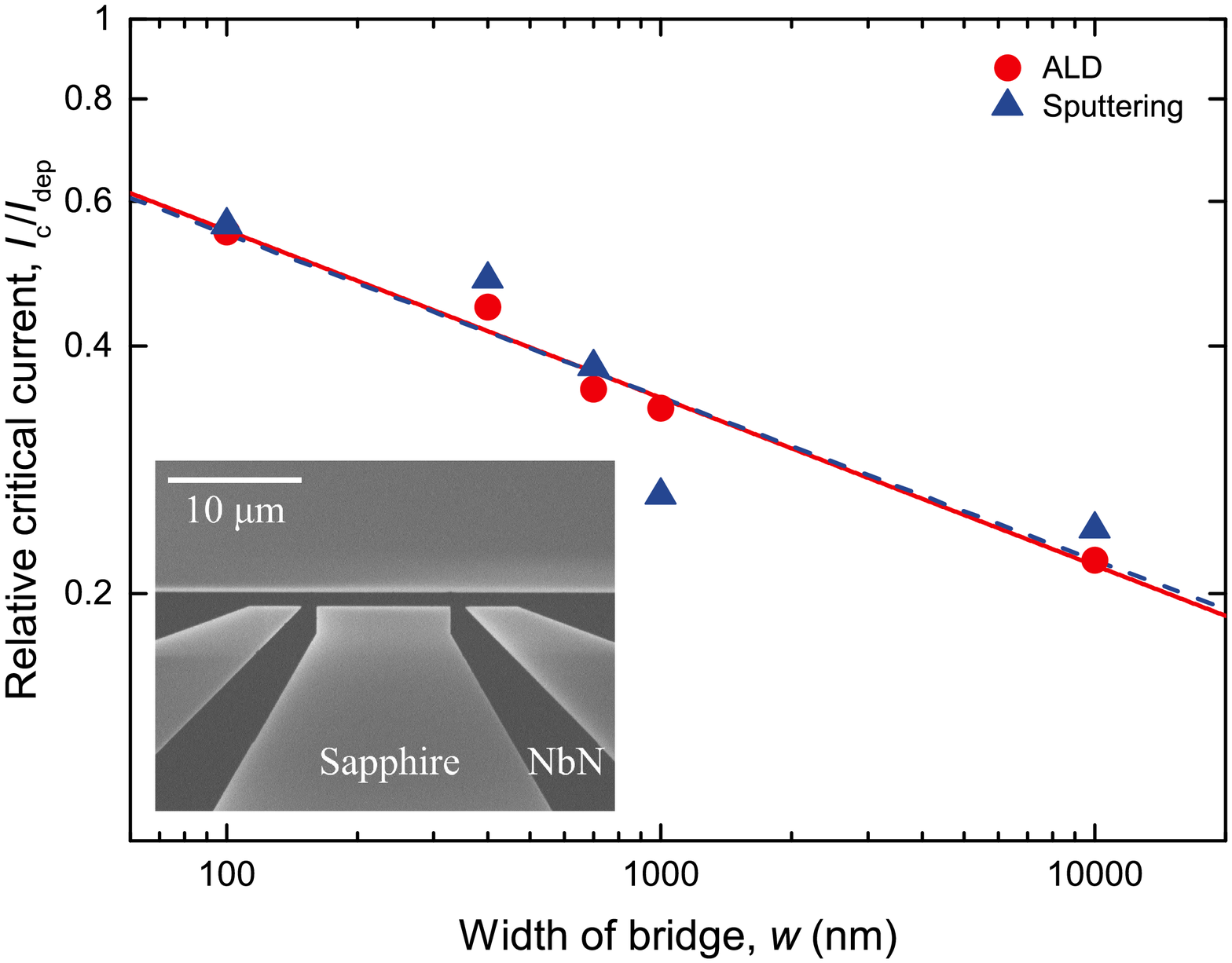}
	\captionof{figure}{Dependence of the relative critical current (ratio of measured critical current $I_\mathrm{c}$ to calculated theoretical depairing current $I_\mathrm{dep}$) on bridge width for different deposition technologies. The film thickness was constant at \SI{10}{nm}. The solid and dashed lines are linear fits to the respective data. (Inset) Exemplary SEM image of the bridge.}
	\label{fig:fig3}
\end{minipage}
\end{figure*}

To define the electron diffusion coefficient $D$ and the coherence length $\xi_\mathrm{GL}$, the upper critical magnetic field $B_\mathrm{c2}$ has been measured as a function of the temperature (see inset in figure 1(d)). From the slope of the linear part of $B_\mathrm{c2}(T)$, the coefficient $D$ was then calculated by
\begin{equation}
	D = - \frac{4 k_\mathrm{B}}{\pi e} \left( \frac{\mathrm{d}B_\mathrm{c2}}{\mathrm{d}T} \right)^{-1}
\end{equation}
with the Boltzmann constant $k_\mathrm{B}$ and the elementary charge $e$ (see equation (9.81) in~\cite{Bartolf:2016}). From the linear part of $B_\mathrm{c2}(T)$, the value $B_\mathrm{c2}(0)$ was calculated by
\begin{equation}
	B_\mathrm{c2}(0) = 0.69\ T_\mathrm{c} \left. \frac{\mathrm{d}B_\mathrm{c2}}{\mathrm{d}T} \right|_{T \lesssim T_\mathrm{c}}
\end{equation}
(see equation~(9.32) in~\cite{Bartolf:2016}, \cite{Werthamer:1966}). The Ginzburg-Landau coherence length at 0~K was estimated by
\begin{equation}
	\xi_\mathrm{GL}(0) = \sqrt{\frac{\Phi_0}{2 \pi B_\mathrm{c2}(0)}}
\end{equation}
with the magnetic flux quantum $\Phi_0 = h/2e$ (see equation~(9.25) in~\cite{Bartolf:2016}).

Compared to reported values for sputtered films, the residual resistivity of ALD films is about two times higher, while the diffusion coefficient is 1.3-2.0 times lower for the same thickness~\cite{Semenov:2009}. Consequently, this should be attended by a similar single-spin density of states at the Fermi-level $N(0) = \left( 2 e^2 \rho_0 D \right)^{-1}$. As a result, the heat capacity of the electronic system $C_\mathrm{e} = \left( 2 \pi ^2 / 3 \right) k_\mathrm{B}^2 N(0) T$ and therefore the ratio $C_\mathrm{e}/C_\mathrm{ph}$ (with the phonon heat capacity $C_\mathrm{ph}$) is similar to sputtered films~\cite{Vodolazov:2017}. A larger $C_\mathrm{e}/C_\mathrm{ph}$ should facilitate the detection of photons because of the larger fraction of the energy which is transferred to the electronic system.

The diffusion coefficient $D$ is another important parameter when evaluating the influence of an absorbed photon on the superconducting order parameter. A small value for $D$ results in short thermalization times $\tau_\mathrm{th}$ in a relatively small hot-spot volume, which improves the detection efficiency~\cite{Vodolazov:2017}. At \SI{10}{nm} thickness, the calculated $D$ for ALD films is almost half the reference value for sputtered films (see figure 1(c))~\cite{Semenov:2009}, which confines the absorbed energy in a smaller volume and, subsequently, leads to a stronger suppression of the order parameter in this hotspot. The diffusion coefficient for NbN is expected to increase with thicker films, as the electron mean free path can be assumed to increase. Above \SI{6}{nm}, the measured $D$ differs from this expected dependence. One reason could be oxygen incorporation into the films as a result of the long ALD deposition times together with oxygen permeation into the deposition chamber. This results in growing defects (niobium oxide) in the deposited film and could lead to a decrease of the diffusion coefficient~\cite{Ziegler:2017}.

In figure 1(d), a decrease of $\xi_\mathrm{GL}$ with thicker films can be observed. This is expected due to the increase of $T_\mathrm{c}$ and the relation $\xi_\mathrm{GL}(0) \propto \sqrt{\xi_0 l}$ where $\xi_0 \propto T_\mathrm{c}^{-1}$ ($\xi_0$ being the BCS coherence length and $l$ the mean free path of electrons, see equation~(9.26) and~(9.28) in~\cite{Bartolf:2016}). The calculated value $\xi_\mathrm{GL}(0) = \SI{4.23}{nm}$ for $d = \SI{5}{nm}$ of ALD-NbN coincides well with a reference value ($\xi_\mathrm{GL}(0) \approx \SI{4.37}{nm}$ for $d = \SI{5.1}{nm}$) of sputtered NbN films of the same thickness on sapphire substrates~\cite{Semenov:2009}.

\subsection{NbN bridges}

NbN bridges have been patterned to investigate the critical current densities $j_\mathrm{c}$. The bridges have a width $w$ between $0.1$ and \SI{10.0}{\micro\meter} and a length $l = 10w$ (see geometry in the inset of figure~3). A reduction of the measurable critical current $I_\mathrm{c}$ due to a non-uniform current distribution across the bridge could be neglected, since $w \ll \Lambda$ (with the Pearl length $\Lambda = \left( 2 \lambda^2 \right) / d$ for thin films with $d \ll \lambda$)~\cite{Pearl:1964,Ilin:2005}.

In figure~2, the dependence of $j_\mathrm{c}$ on the normalized temperature $T/T_\mathrm{c}$ is shown. The experimental values have been compared to the temperature dependence of the depairing critical current density $j_\mathrm{dep}(T)$ according to the derivation of Kupriyanov and Lukichev~\cite{Kupriyanov:1980}
\begin{equation}
	j_\mathrm{dep}(T) = j_\mathrm{dep,GL}(0) \left( 1 - \frac{T}{T_\mathrm{c}} \right) ^{\frac{3}{2}} \mathrm{KL}(T),
\end{equation}
with the formal Ginzburg-Landau zero-temperature depairing current
\begin{equation}
	j_\mathrm{dep,GL}(0) \approx 1.454\ \beta_0^2 \frac{\left( k_\mathrm{B} T_\mathrm{c} \right) ^{\frac{3}{2}}}{e \rho_0 \sqrt{D \hbar}}
\end{equation}
and the correction factor $\mathrm{KL}(T)$ for the dirty limit (see curve~7, figure~1 in~\cite{Kupriyanov:1980}).

In figure 3, the ratio $I_\mathrm{c}/I_\mathrm{dep}$ is plotted for both ALD and sputtered NbN bridges. Samples of both types of deposition were patterned in parallel.

The temperature dependencies of $j_\mathrm{c}$ for different widths coincide near $T_\mathrm{c}$ as expected from the Ginzburg-Landau theory (figure~2). For smaller ratios $T/T_\mathrm{c}$, the penetration and dissipative movement of vortices across the bridge suppress the measurable value $j_\mathrm{c}$ for larger bridge widths.

In addition to the theoretical $j_\mathrm{dep}(T)$ calculated by equation~(4) and~(5) (solid line, figure 2), a fit to the experimental data was also plotted (dashed lines). For these curves, we used $j_\mathrm{dep,GL}(0)$ as a fitting parameter. As it is seen, only the data for $w = \SI{0.1}{\micro\meter}$ can be well fitted by equation~(4), which could indicate that the ratio $j_\mathrm{c}/j_\mathrm{dep}$ is only reduced by constrictions. In this case, the fit estimates the relative cross-section of the superconducting core.

Although the absolute values of $j_\mathrm{c}$ for the ALD-NbN bridges were significantly lower than the ones for sputtered NbN ($9$ compared to \SI{23}{MA/cm^2} for $w = \SI{0.1}{\micro\meter}$), due to a higher $\rho_0$ and a lower $T_\mathrm{c}$, the calculated ratios of $I_\mathrm{c}$ to $I_\mathrm{dep}$ are comparable, reaching almost $0.6$ at the minimal width of \SI{0.1}{\micro\meter} (figure~3). According to~\cite{Korzh:2018,Vodolazov:2017}, a higher ratio $I_\mathrm{c}/I_\mathrm{dep}$ should lead to higher internal detection efficiencies for lower-energy photons and to a smaller intrinsic timing jitter.

The obtained ratios are still below one, which is probably due to both material-inherent non-uniformities and fabrication imperfections of the nanowire geometry. Moreover, for the calculations we use the nominal width, determined by SEM, without taking the edge roughness into account. Also, damaged normal-conducting bands around the superconducting core were neglected. These are expected due to film oxidation, the formation of an intermediate layer between the substrate and the film and damaged edges during etching, as described in~\cite{Fominov:2001}.

\section{Nanowire detectors}

We fabricated straight nanowires with a width $w = \SI{100}{nm}$ and a length $l = \SI{5}{\micro\meter}$ for the investigation of detector properties. Detectors of both \SI{5}{nm} and \SI{10}{nm} film thickness have been characterized. To minimize current-crowding~\cite{Hortensius:2012}, the nanowire is connected by short tapers at both ends. Since the low kinetic inductance of these short nanowires would lead to latching, a \SI{1}{\micro\meter}-wide, \SI{1}{mm}-long inductor in series to the nanowire was patterned, adding a kinetic inductance of about $L_\mathrm{kin} \approx \SI{46}{nH}$ for $d = \SI{10}{nm}$ ($L_\mathrm{kin} = \mu_0 \lambda^2 l / w d$, see e.g. equation~(9.45) in~\cite{Bartolf:2016}). The detectors have critical temperatures in the range of $9$ to \SI{11}{K} and critical currents of $15$ and \SI{43}{\micro\ampere} for $d = \SI{5}{nm}$ and \SI{10}{nm}, respectively.

\subsection{Experimental setup}

\begin{figure*}[t]
	\centering
	\subfloat{\includegraphics[width=0.49\textwidth]{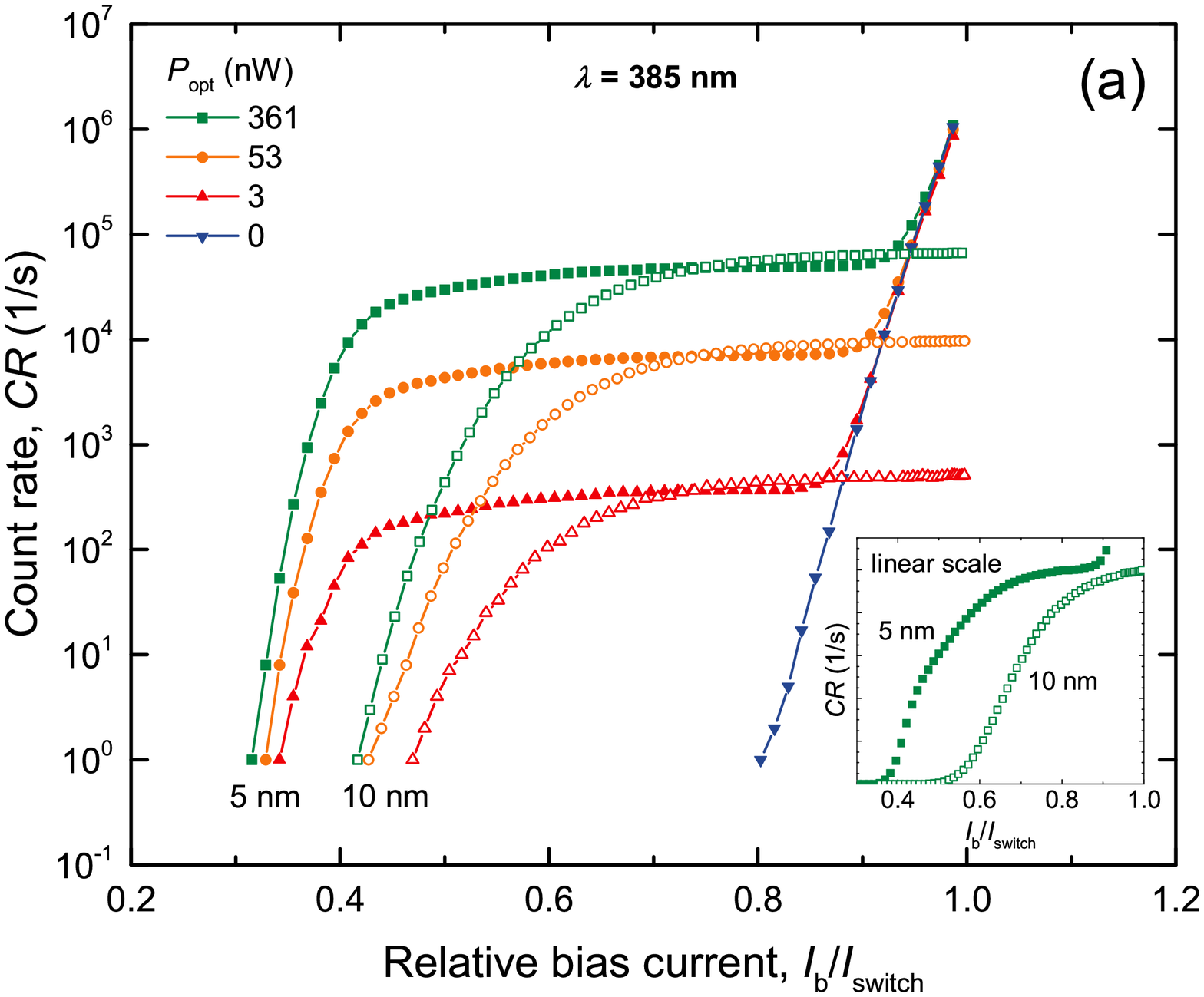}} \hfill
	\subfloat{\includegraphics[width=0.49\textwidth]{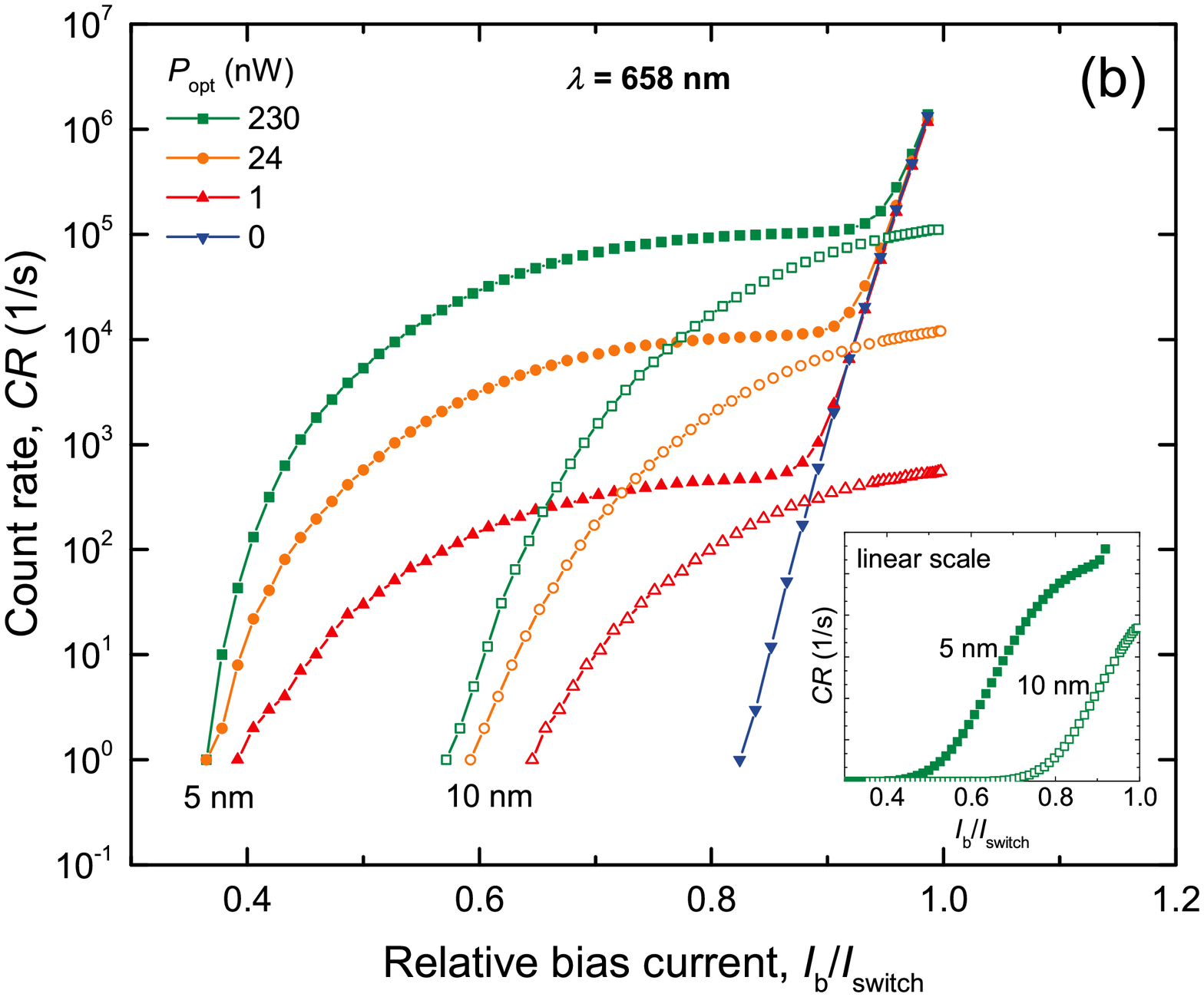}}
	\caption{Dependence of the count rate $CR$ on the relative bias current $I_\mathrm{b}/I_\mathrm{switch}$ for illumination with photons at a wavelength of \SI{385}{nm} (a) and \SI{658}{nm} (b) and different optical powers. (Insets) Plots on the linear scale for the highest power level.}
	\label{fig:fig4}
\end{figure*}

In order to characterize the detectors optically, the samples were mounted in a dipstick cryostat equipped with an optical fiber, RF, and DC lines. The dipstick was cooled down in lHe to \SI{4.2}{K}. For the detector’s operation, a low-noise voltage source with RC-filters connected to a room-temperature bias-tee was used. The samples were illuminated from the top by different light sources. For measurements dependent on the bias current and the optical power, the output of an LED at \SI{385}{nm} and a laser diode at \SI{658}{nm} was coupled on to the detector via a multimode fiber, which is specified from $400$ to \SI{2400}{nm}. For spectral measurements, a halogen lamp and a monochromator for the wavelength range between $400$ and \SI{2000}{nm} were used. In order to ensure a uniform illumination of the detectors, the samples were kept at a distance of \SIrange[range-phrase = --, range-units = single]{4}{5}{\milli\meter} from the fiber tip resulting in an illuminated area with a radius of about \SI{1}{\milli\meter}.

All of the light sources were calibrated using a power meter equipped with Si- and InGaAs-based photodiodes for the wavelength range $350$ to \SI{1100}{nm} and $800$ to \SI{1700}{nm} (Thorlabs S150C and S154C), respectively. The calibration was done at the detector plane at the bottom of the dipstick in order to take into account any wavelength-dependent loss contribution by the vacuum fiber feedtrough, the fiber itself and the output of the stripped fiber.

The count rate was measured by a pulse counter (up to \SI{300}{MHz} count rate) with a two-stage room-temperature preamplifier connected to the RF output of the bias-tee. The resulting pulse height was in the range of $200$ to \SI{900}{\milli\volt}, depending on the critical current of the sample. The pulse and jitter of the detector were monitored using a real-time oscilloscope of type Agilent Technologies DSA-X 93204A.

The output optical power from the monochromator was strongly dependent on the wavelength. In order to keep the detector in the same detection regime during spectral measurements, the detection count rate was kept at values below \SI[retain-unity-mantissa = false]{1e4}{\per\second} by manually adjusting the input slit width of the monochromator accordingly. The measured data was then stitched by means of overlapping data points.

\subsection{Results}

The bias-dependent count rates for different film thicknesses, wavelengths, and optical input powers are shown in figure~4. For $\lambda = \SI{385}{nm}$ (figure 4(a)), all measured curves exhibit saturating count rates, which is attributed to near-unity internal detection efficiency for high bias currents~\cite{Vodolazov:2014}. For lower bias currents, the count rate decreases exponentially. In this, so called probabilistic or fluctuation-assisted detection regime, a detection event can only occur with the help of fluctuations. As indicated by the prolonged plateau in $CR(I_\mathrm{b})$, the thinner sample shows a deterministic regime over a wider bias range. At $\lambda = \SI{658}{nm}$ (figure 4(b)), the plateaus are less pronounced (\SI{5}{nm}) or not existing anymore (\SI{10}{nm}). To better illustrate this, insets with curves on the linear scale are included in figure~4.

For the same optical input power, the level of saturation is largely independent of the sample thickness, further emphasizing the near-unity internal efficiency.

By blocking the feed-through of the fiber, the dark-count rate ($DCR$) was obtained. We measured an $DCR$ exponentially growing with the bias current for $d = \SI{5}{nm}$ and $I_\mathrm{b} > 0.8\,I_\mathrm{switch}$, whereas the \SI{10}{nm} sample did not exhibit an increasing $DCR$ above a noise level~$\leq \SI{3}{\per\second}$ up to $I_\mathrm{switch}$.

\begin{figure*}[t]
\centering
\begin{minipage}[t]{.485\textwidth}
	\centering
	\includegraphics[width=\linewidth]{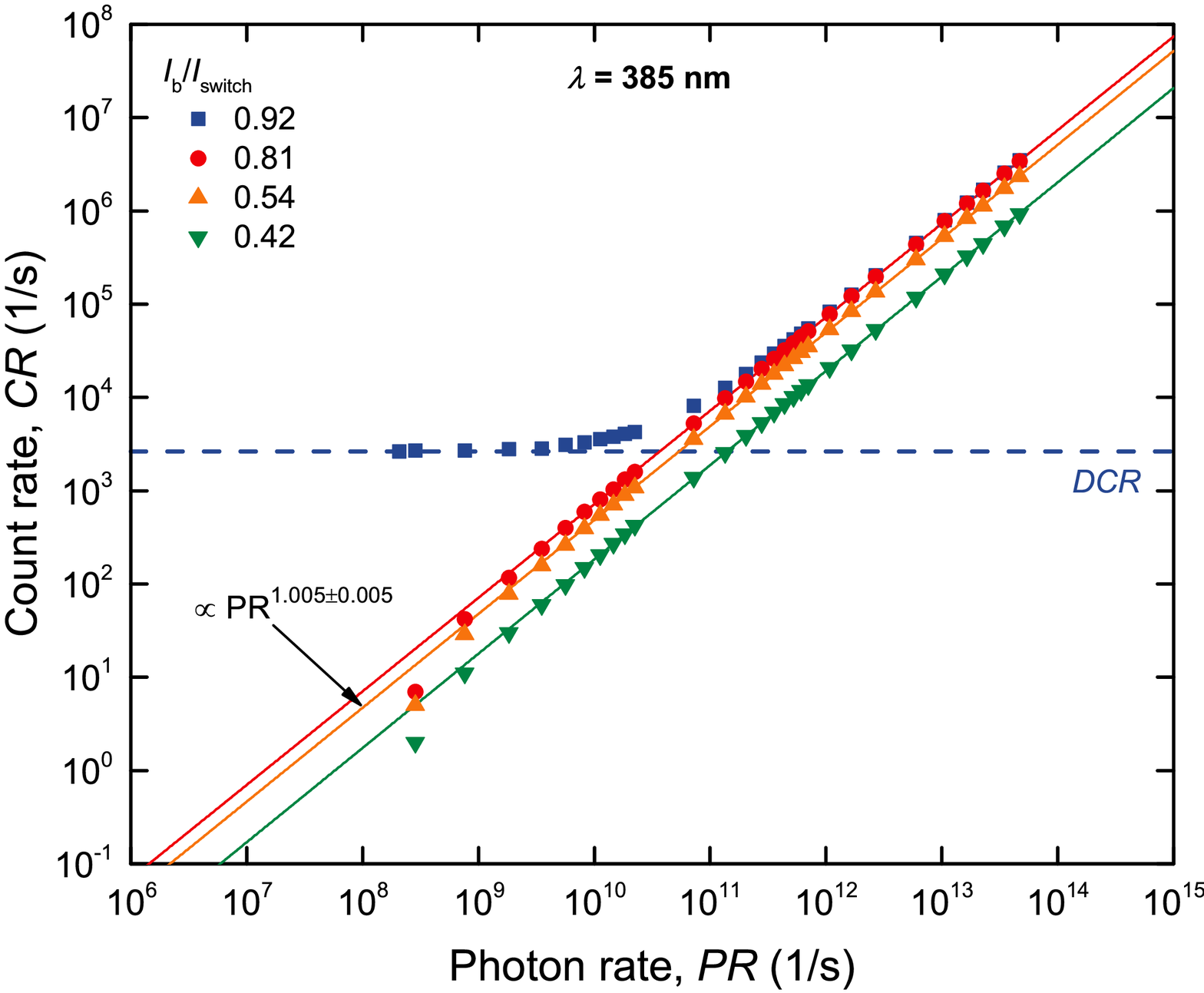}
	\captionof{figure}{Dependence of the count rate $CR$ on incoming photon rate $PR$ for different bias currents for a detector with thickness $d = \SI{5}{nm}$ at $\lambda = \SI{385}{nm}$. The solid lines are linear fits to the respective data. The dashed line indicates the level of dark counts at $0.92\,I_\mathrm{switch}$.}
	\label{fig:fig5}
\end{minipage}\hfill%
\begin{minipage}[t]{.485\textwidth}
	\centering
	\includegraphics[width=\linewidth]{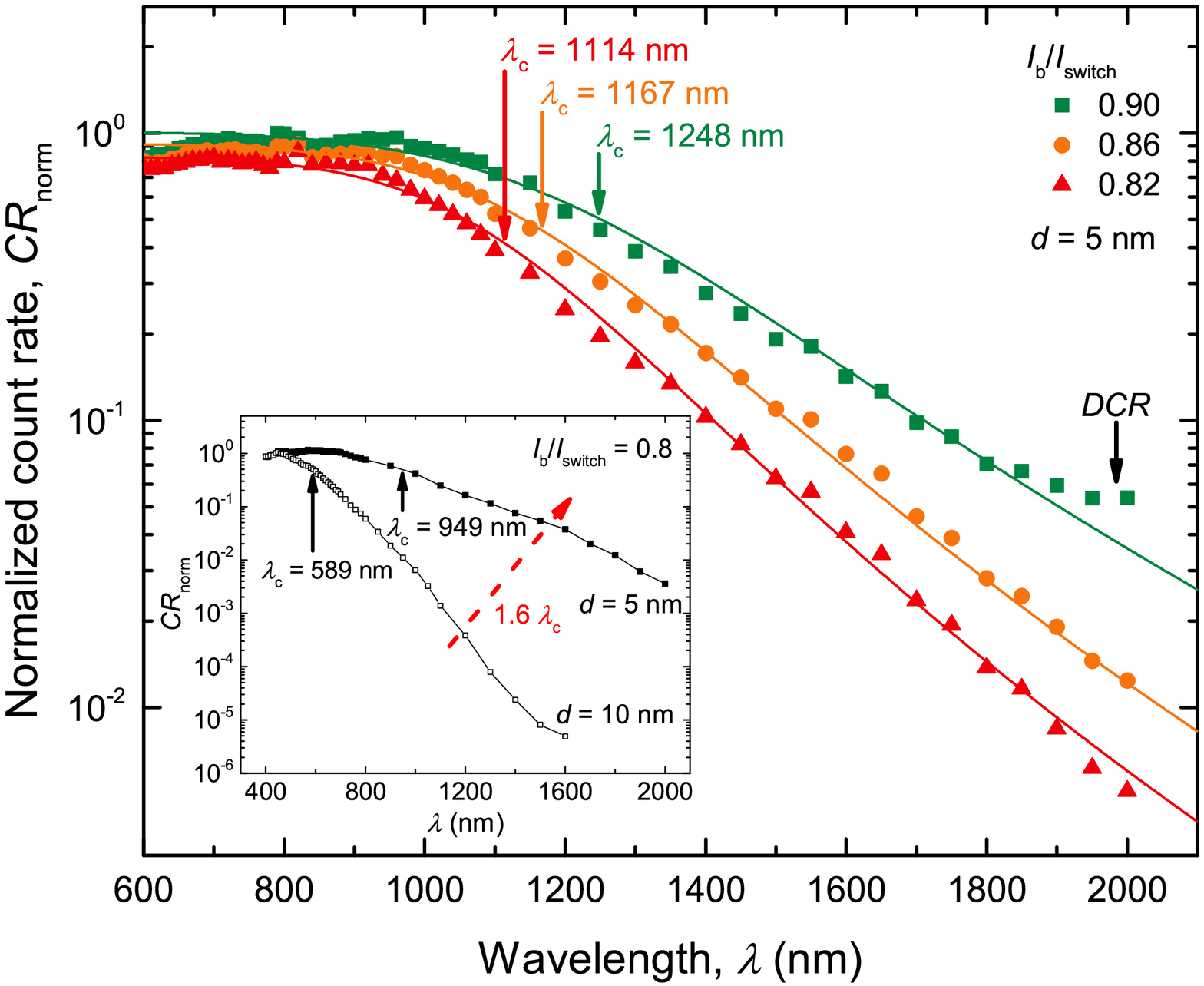}
	\captionof{figure}{Dependence of the normalized count rate $CR_\mathrm{norm}$ on the illumination wavelength $\lambda$ of a \SI{5}{nm}-thick detector. The solid lines are fits with equation~(6). (Inset) Comparison of the normalized count rate $CR_\mathrm{norm}$ of \num{5} and \SI{10}{nm}-thick detectors with the same $I_\mathrm{b}/I_\mathrm{switch}$ ratio of \num{0.8}.}
	\label{fig:fig6}
\end{minipage}
\end{figure*}

As illustrated in figure~5, the count rate increases with the incoming photon rate $PR$ as $CR \propto PR^n$ with an exponent $n$ of about~\num[separate-uncertainty]{1.005 +- 0.005}. This applies to all measured $I_\mathrm{b}/I_\mathrm{switch}$-ratios corresponding to different bias regimes. The linear dependence of $CR \propto PR$ over many orders of magnitude is a strong indication of single-photon detection~\cite{Goltsman:2001,Korneeva:2016}. A saturation of the count rate $CR_\mathrm{max}$ could not be reached because of the weak optical coupling and the subsequent heating of the sample, suppressing $I_\mathrm{switch}$ and therefore limiting $I_\mathrm{b}$. For $I_\mathrm{b} > 0.81\,I_\mathrm{switch}$ and low photon rates, the count rate converges to $DCR$. The divergence of the data points with respect to the linear progression at lower $PR$ is due to the calibration precision of the power meter (about \SI{100}{pW} or $PR \approx \SI{2e8}{\per\second}$ at $\lambda = \SI{385}{nm}$). Similar results were obtained for $d = \SI{10}{nm}$ and $\lambda = \SI{658}{nm}$.

In figure~6, the normalized count rate as a function of the wavelength of the incident photons from $600$ to \SI{2000}{nm} is shown. The spectral count rates have a plateau at lower wavelengths and drop monotonically above the cut-off wavelength.

The data were fitted by
\begin{equation}
	SDE \propto \left( 1 + \left( \frac{\lambda}{\lambda_\mathrm{c}} \right) ^p \right) ^{-1}
\end{equation}
with a cut-off wavelength $\lambda_\mathrm{c}$ and an exponent $p$ indicating the decay slope above $\lambda_\mathrm{c}$~\cite{Charaev:2017a}. The highest cut-off wavelength obtained by this fit was about \SI{1248}{nm} for $d = \SI{5}{nm}$ ($I_\mathrm{b} = 0.90\,I_\mathrm{switch}$) and \SI{878}{nm} for $d = \SI{10}{nm}$ ($I_\mathrm{b} = 0.98\,I_\mathrm{switch}$).

Here, we should note different definitions of the cut-off wavelength. Besides the fit used herein, the intersection of two straight lines is sometimes used, too~\cite{Charaev:2017a}. The latter method, in general, leads to slightly lower values of $\lambda_\mathrm{c}$.

The estimated timing jitter of the ALD SNSPDs lies well below \SI{28}{ps}, which is the limit of the used experimental setup due to intermodal dispersion in the multimode fiber.

\subsection{Discussion}

According to existing detection models for SNSPDs, events of both photon counts in the probabilistic regime and dark counts could be explained by fluctuations, e.g. vortex hopping~\cite{Bulaevskii:2012,Zotova:2014}, vortex-antivortex-pair (VAP) unbinding~\cite{Zotova:2012,Engel:2015,Vodolazov:2015,Engel:2005,Kitaygorsky:2007,Yamashita:2011}, or Fano fluctuations~\cite{Kozorezov:2017}.

In the framework of these models, the observed low $DCR$ for thicker ALD-NbN films might be explained by the smaller Pearl length, which, in turn, increases the edge barrier for vortex penetration~(see equation~(20) and (21) in~\cite{Bartolf:2010}). In addition, the larger volume with the same area of the film-substrate interface of the nanowire should result in smaller thermal fluctuations.

As a second mechanism, dark-counts due to the unbinding of VAPs should be suppressed in \SI{10}{nm}-thick nanowires, since the film thickness is more than twice the coherence length $\xi_\mathrm{GL}(\SI{4.2}{K}) \approx \SI{4.45}{nm}$ and the probability of VAP excitations is therefore lower.

It can be observed, that $\lambda_\mathrm{c}$ increases with rising bias currents as it is at least qualitatively described by various detection models~\cite{Semenov:2005,Vodolazov:2014,Vodolazov:2015,Maingault:2010}. According to these models, the superconducting state becomes unstable after the absorption of lower energy photons as the current is increased. This can be explained by the reduced edge-barrier for single vortices and facilitated unbinding of VAPs nucleated inside the hotspot~\cite{Vodolazov:2017,Vodolazov:2015,Bartolf:2010}. Also, larger portions of the nanowire across the width contribute to the photon detection for increasing bias currents~\cite{Renema:2015}.

Besides that, the detector layout is beneficial for detecting lower-energy photons, since without bends, the single straight nanowires suffer less from current-crowding and, due to its short length, have a reduced probability of material non-uniformities.

As illustrated by the inset of figure~6, we get an increase of the cut-off wavelength $\lambda_\mathrm{c}$ by a factor of $1.6$ for the same ratio $I_\mathrm{b}/I_\mathrm{switch}$ by reducing the thickness from $10$ to \SI{5}{nm}. This is in line with the expected proportionality $\lambda_\mathrm{c} \propto d^{-1}$, since less demand is made for the ratio $I_\mathrm{b}/I_\mathrm{dep}$ for detectors with smaller cross-sections~\cite{Semenov:2005,Vodolazov:2017}. A change of $\lambda_\mathrm{c}$ by a factor smaller than $2.0$ is, at least partly, due to the fact, that material parameters such as $\Delta$, $D$ and $\tau_\mathrm{th}$ also change with the thickness (see equation~(4) in~\cite{Semenov:2005}). Furthermore, the actual superconducting core of the strip is, relative to its nominal cross-section, smaller for $5$ than for \SI{10}{nm}, as the damaged bands around the superconducting core are expected to remain constant in size.

\section{Conclusion}

Using the ALD technique, we deposited superconducting NbN thin films on sapphire substrates. With a small diffusion coefficient ($D = \SI{0.32}{\centi\meter\squared\per\second}$ for $d = \SI{10}{nm}$) and similar values for the ratios $C_\mathrm{e}/C_\mathrm{ph}$ and $I_\mathrm{c}/I_\mathrm{dep}$ ($\approx \num{0.56}$) compared to sputtered NbN, these highly disordered films exhibit well-suited properties for the use in nanowire detectors.

SNSPDs based on this atomic layer-deposited NbN have been successfully demonstrated. These detectors show saturated $CR(\lambda)$ characteristics and a cut-off wavelength of about \SI{1.25}{\micro\meter} for a relatively large cross-section of \SI{5x100}{nm}. Decreasing the nanowire width should lead to a further increase of the cut-off wavelength~\cite{Marsili:2012}.

We expect, that the ALD technology will gain in importance in the future, since it offers greater flexibility regarding substrate materials and wafer size in comparison to existing deposition techniques. In this regard, the homogeneity with respect to both thickness and stoichiometry over large areas should be characterized. The high degree of film uniformity is important for reproducibility and yield of SNSPDs when scaling up to arrays.

\ack{The authors acknowledge S. Goerke, K. Pippardt, and U.~Hübner for support with the thin film fabrication and S. Linzen for support with the development of the deposition technology.

Part of this work was funded by the Federal Ministry of Education and Research under contract no. 13N13445.}


\end{document}